\documentclass[prb,twocolumn,superscriptaddress, showpacs,floatfix]{revtex4-1}  
\usepackage{hyperref}
\usepackage{amssymb}
\usepackage{amsmath}
\usepackage{float}
\usepackage[usenames]{color}
\usepackage{graphicx}

\begin{document}

\title{Systematically improvable multi-scale solver for correlated electron systems}

\author{Alexei A. Kananenka}
\affiliation{Department of Chemistry, University of Michigan, Ann Arbor, Michigan, 48109, USA}
\author{Emanuel Gull}
\affiliation{Department of Physics, University of Michigan, Ann Arbor, Michigan, 48109, USA}
\author{Dominika Zgid}
\affiliation{Department of Chemistry, University of Michigan, Ann Arbor, Michigan, 48109, USA}

\begin{abstract}
The development of  numerical methods capable of simulating realistic materials with strongly correlated electrons, with controllable errors,
 is a central challenge in quantum many-body physics. Here we describe a framework for a general multi-scale method based on embedding a self-energy of a strongly correlated subsystem into a self-energy generated by a method able to treat large weakly correlated systems approximately.
As an example, we present the embedding of an exact diagonalization self-energy into a self-energy generated from self-consistent second order perturbation theory. 
Using a quantum impurity model, generated from a cluster dynamical mean field approximation to the 2D Hubbard model, as a benchmark,
we illustrate that our method allows us to obtain accurate results at a fraction of the cost of typical Monte Carlo calculations.
We test the method in multiple regimes of interaction strengths and doping of the model. 
The general embedding framework we present avoids difficulties such as  double counting corrections, frequency dependent interactions, or vertex functions. As it is solely formulated at the level of the single-particle Green's function, it  provides a promising route for the simulation of realistic materials that are currently difficult to study with other methods.
\end{abstract}


\maketitle

\section{Introduction and General Framework}

The theoretical description of strongly correlated materials has proven to be challenging, mainly because many of their interesting properties are caused by the interplay of subtle electronic correlation effects on low energy scales. Since simultaneous treatment of both strong and weak correlations is of major importance for the quantitative description of these systems, two main conceptual approaches are used: the reduction to a few `relevant' degrees of freedom or essential orbitals around the Fermi level and the subsequent construction of a model system or, alternatively, the treatment of the entire system using methods which significantly approximate correlation effects.

The first approach, with methods including exact diagonalization (ED)~\cite{Davidson197587,Lin93,0953-8984-24-5-053201,PhysRevB.76.245116} and its variants,\cite{Zgid11,Zgid12} density matrix renormalization group (DMRG),\cite{PhysRevLett.69.2863} dynamical mean field theory (DMFT)\cite{Georges96,Maier05} and lattice quantum Monte Carlo (QMC)\cite{BSS81} applied to model Hamiltonians, can yield very precise results for model systems. When applied to realistic systems, its main uncertainties and possible sources of errors lie in the construction of the parameters of the effective model.

The second approach, which includes implementations of the density functional theory (DFT),\cite{PhysRev.136.B864,PhysRev.136.B864,DFT_YP} Hartree Fock (HF), GW,\cite{HedinGWpra1965} the random phase approximation (RPA),\cite{PhysRev.82.625}  M\o ller-Plesset second order perturbation theory (MP2),\cite{PhysRev.46.618} GF2~\cite{:/content/aip/journal/jcp/140/24/10.1063/1.4884951,Dahlenjcp2005} or QMC,\cite{RevModPhys.73.33} avoids constructing an effective model by treating the full Hamiltonian with all orbitals and interactions, but relies on potentially severe approximations to the electronic correlations.

Multi-scale methods for extended systems combining the best aspects of both approaches, {\it e.g.} by solving the system using DFT or GW and using the result to construct a model system, have been implemented as the GW+DMFT\cite{Biermann03,SilkeGW04,Karlsson05,Kotliar06,Held07,Tomczak12,Casula12,Casula12b,Ayral12,Taranto13,Sakuma13,Hansmann13,Ayral13,Biermann14}  and DFT+DMFT\cite{Kotliar06,Held07} method.

Constructing a robust multi-scale method is a formidable problem and an active field of research. 
First, different energy scales have to be defined and a set of strongly correlated orbitals requiring a higher level treatment has to be chosen. Second, the non-local Coulomb interactions present in realistic materials have to be included by a suitable choice of `screened' interactions. Third, correlations in the weakly correlated orbitals  should not be completely neglected but rather be treated perturbatively, if a quantitative material-dependent description is desired.

In this paper, we present a general framework for a multi-scale algorithm 
in which a self-energy describing strongly correlated orbitals is self-consistently embedded  into the a self-energy obtained from a method able to treat long-range interaction and correlation effects. 
We call this general framework `self-energy embedding theory' (SEET). 
A mathematically rigorous procedure for identifying the strongly correlated orbitals and systematically increasing the accuracy of the treatment of the weakly correlated orbitals is an integral part of our procedure.  
The strongly and the weakly correlated subspaces are treated using different methods. As an example, we will choose exact diagonalization (ED) to yield the self-energy for the strongly correlated orbitals and the self-consistent second order Green's function method (GF2)\cite{:/content/aip/journal/jcp/140/24/10.1063/1.4884951} for the weakly correlated part, and call the resulting algorithm SEET(ED-in-GF2). However,we note that our scheme is general and independent of the algorithms used to treat the weakly and strongly treated subspaces: ED could be replaced by, {\it e.g.}, CT-QMC, while GF2 could be replaced with FLEX, GW, or the Parquet method.  

 Here we illustrate SEET(ED-in-GF2) and calibrate it using impurity problems, since a method capable of treating multi-scale problems such as realistic materials should also yield accurate results for model systems.  Impurity problems have a continuous dispersion but only a finite number of interaction terms. Nevertheless, they exhibit strongly and weakly correlated regimes and a number of phases and phase transitions that are very well understood and for which numerically exact comparison algorithms exist.\cite{RevModPhys.83.349} The restriction to these models allows us to provide stringent tests on the accuracy of SEET in a well controlled test environment. In the future, we aim to apply SEET to realistic materials, where it has the potential to become a method complementary to GW+DMFT or DFT+DMFT.

In order to generate a wide range of correlated phases, we generate our impurity parameters from a four-site dynamical cluster approximation (DCA)\cite{Hettler00,Maier05} to the 2D Hubbard model, {i.e.} test our method as a `DCA impurity solver', so that our results can be compared against numerically exact continuous time CT-QMC data.\cite{Gull08_ctaux,PhysRevB.83.075122} We emphasize that while we have developed a numerically efficient impurity solver, we do not envisage this as the main use of SEET, and we mainly resort to impurity models for the sake of generating comparisons to reliable known results.

SEET in the ED-in-GF2 variant is computationally affordable, as GF2 scales as $O(N^5)$, where $N$ is the number of orbitals in a unit cell. Additionally, SEET is amenable to parallelization on large machines. The scaling of GF2  can further be reduced to $O(N^4)$ by employing density fitted integrals,\cite{:/content/aip/journal/jcp/140/24/10.1063/1.4884951} and the strongly correlated orbitals can be treated by ED as pairs, further reducing the numerical cost. Consequently, large systems containing many unit cells or $k$-points containing multiple orbitals can be treated simultaneously, providing non-local effects and momentum dependence.

In Section \ref{sec:gfpdmft_method}, we introduce SEET(ED-in-GF2). Sec.~\ref{sec:results} shows results for our test model, and Sec.~\ref{sec:conclusions} contains conclusions of our work.
\section{The SEET(ED-in-GF2) method applied to an impurity model}\label{sec:gfpdmft_method}

We consider an impurity problem with $N$ impurity orbitals $a_i$
coupled to an infinitely many bath orbitals $c_\mu$ described by a general Hamiltonian
\begin{equation}\label{ham}
\begin{split}
\hat{H}=\sum_{ij} t_{ij}a^{\dagger}_{i}a_{j}+\sum_{ijkl}U_{ijkl}a^{\dagger}_{i}a^{\dagger}_{j}a_{l}a_{k}+\\
+\sum_{i\lambda}V_{i\lambda}a^{\dagger}_{i}c_{\lambda} +\sum_{\lambda}\epsilon_{\lambda}c^{\dagger}_{\lambda}c_{\lambda}+h.c.,
\end{split}
\end{equation}
where $t$ and $U$ are material specific one- and two-body operators, $V$ is the hybridization strength, and $\epsilon_\lambda$ is the $c$-electron dispersion. The single-particle properties of this Hamiltonian are described by
a non-interacting Matsubara Green's function for $a$-electrons 
\begin{equation}
G_{0}(i\omega)=[i\omega+\mu - t -\Delta(i\omega)]^{-1}, \label{G0}
\end{equation}
with $\Delta(i\omega)$ encapsulating the properties of the $c$-electrons and $\mu$ being the chemical potential. In SEET(ED-in-GF2), we obtain the interacting Green's function $G^{GF2}$ of this $N$-orbital impurity problem iteratively, starting from $G^{GF2}=G_0$, by self-consistent second order perturbation theory (GF2),\cite{:/content/aip/journal/jcp/140/24/10.1063/1.4884951,Dahlenjcp2005}
\begin{equation}\label{GF2}
\begin{split}
[\Sigma^{GF2}(\tau)]_{ij}=-\underset{klmnpq}{\sum}[G^{GF2}(\tau)]_{kl}[G^{GF2}(\tau)]_{mn}\times\\
\times[G^{GF2}(-\tau)]_{pq}U_{iqmk}\bigr(2U_{lnpj}-U_{nlpj}\bigr)
\end{split}
\end{equation}
and the corresponding GF2 Green's function 
\begin{equation}
G^{GF2}(i\omega)=\big[[G^{GF2}_{0}(i\omega)]^{-1}-\Sigma^{GF2}(i\omega)\big]^{-1}. 
\end{equation}

\begin{figure}
\centering
\includegraphics[width=1.0\columnwidth]{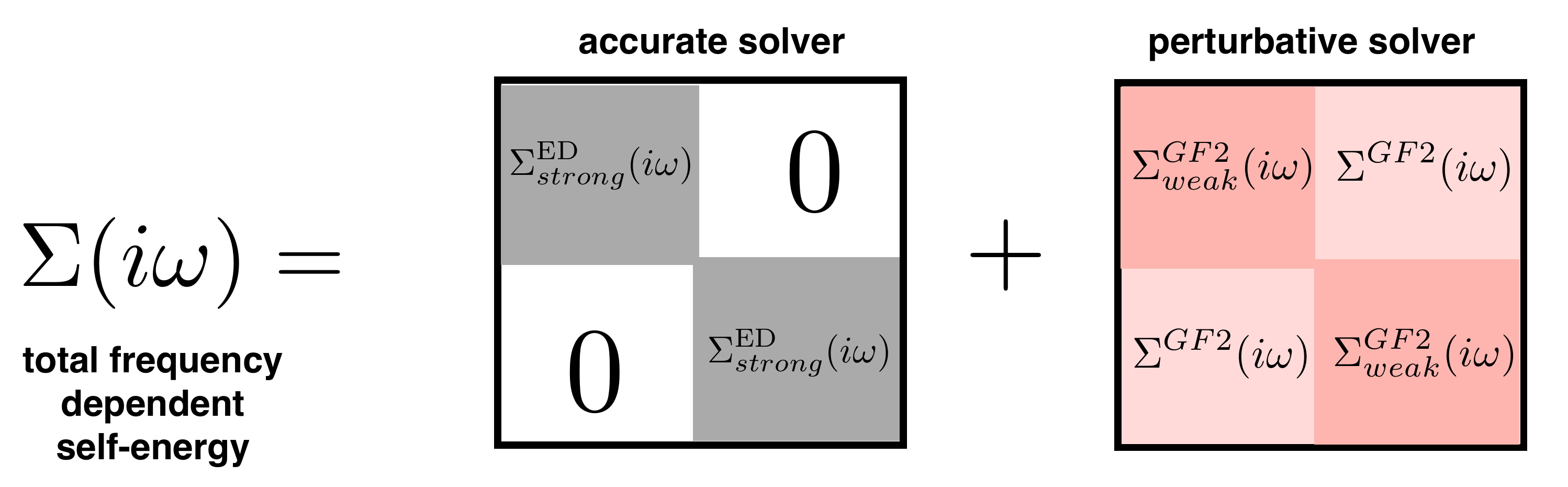}
\caption{\label{se_scheme} The total self-energy in natural orbital basis produced in the SEET with ED-in-GF2 scheme. }
\end{figure}

Note that GF2 can be solved self-consistently and includes an exchange diagram important for describing systems with a localized electronic density, but does not include higher order RPA-like diagrams that are present, {\it e.g.}, in GW.
We then evaluate the one-body density matrix using the converged GF2 Green's function and choose a set of $n<N$ orbitals corresponding to eigenvalues of the one-body density matrix which are significantly different from 0 or 2. 
These $n$ orbitals, which we will call `strongly correlated', are used to build an $n$-orbital impurity problem 
which is then solved with a method more accurate than GF2 to compute a self-energy. Here, we use ED\cite{Caffarel94} to solve this impurity problem. The resulting ED self-energy is used to modify the GF2 self-energy and to obtain the total self-energy in the natural orbital basis as
\begin{equation}\label{se1}
[\Sigma]_{ij}=[\Sigma^{ED}_{strong}]_{\mu\nu}+[\Sigma^{GF2}]_{ij}-[\Sigma^{GF2}_{strong}]_{\mu\nu}. 
\end{equation}
The indices $i$ and $j$ run over all $N$ orbitals, while $\mu$ and $\nu$ run only over the $n$ strongly correlated orbitals. The total self-energy is schematically illustrated in Fig.~\ref{se_scheme}.

As the $n$ correlated orbitals are chosen in the eigenbasis of the one-body density matrix, a transformation of the one-body and two-body integrals in this $n$-orbital subspace to the eigenbasis is necessary. Note, that even for  cases where model Hamiltonians with simplified (e.g. local or density-density) interaction structures are studied, this transformation generates general interactions $U_{ijkl}$.  
This $n$-orbital impurity problem with non-local interaction $U_{ijkl}$ is then treated by the ED solver requiring an additional bath discretization step which may introduce fitting errors. We emphasize that these are small for the cases studied here and that, in principle, any solver suitable to describe strong correlations and able to treat general multi-orbital interactions can be employed, including QMC solvers based on the hybridization expansion \cite{Werner06Kondo} which do not require a bath discretization step.

The ED-in-GF2 procedure is iterated, and the GF2 calculation updates $[\Sigma^{GF2}]_{ij}$ for the $N$ orbitals since $i,j=1,\dots,N$ where
\begin{equation}
[\Sigma^{GF2}_{weak}]_{\mu\nu}=[\Sigma^{GF2}]_{\mu\nu}-[\Sigma^{GF2}_{strong}]_{\mu\nu}
\end{equation}
is responsible for removal of diagrams later included at the ED level. Subsequent ED calculation updates the strongly part of self-energy $[\Sigma^{ED}_{strong}]_{\mu\nu}$. The iterative updates stop when the total self-energy in Eq.~\ref{se1} is converged to a predefined accuracy. We present a detailed algorithmic description of this framework in the supplementary material. 

The algorithm, both in its general form and in the ED-in-GF2 variant, is based on a diagrammatic formulation in which a `double counting'\cite{Karolak201011} problem does not appear. The single-particle formulation avoids vertex functions, which are often difficult to handle, and is based on static (or frequency-independent) interactions.\cite{PhysRevB.70.195104,PhysRevB.86.165105,PhysRevB.87.165118}

\section{Results}\label{sec:results}

\begin{figure}
\centering
\includegraphics[width=1.0\columnwidth]{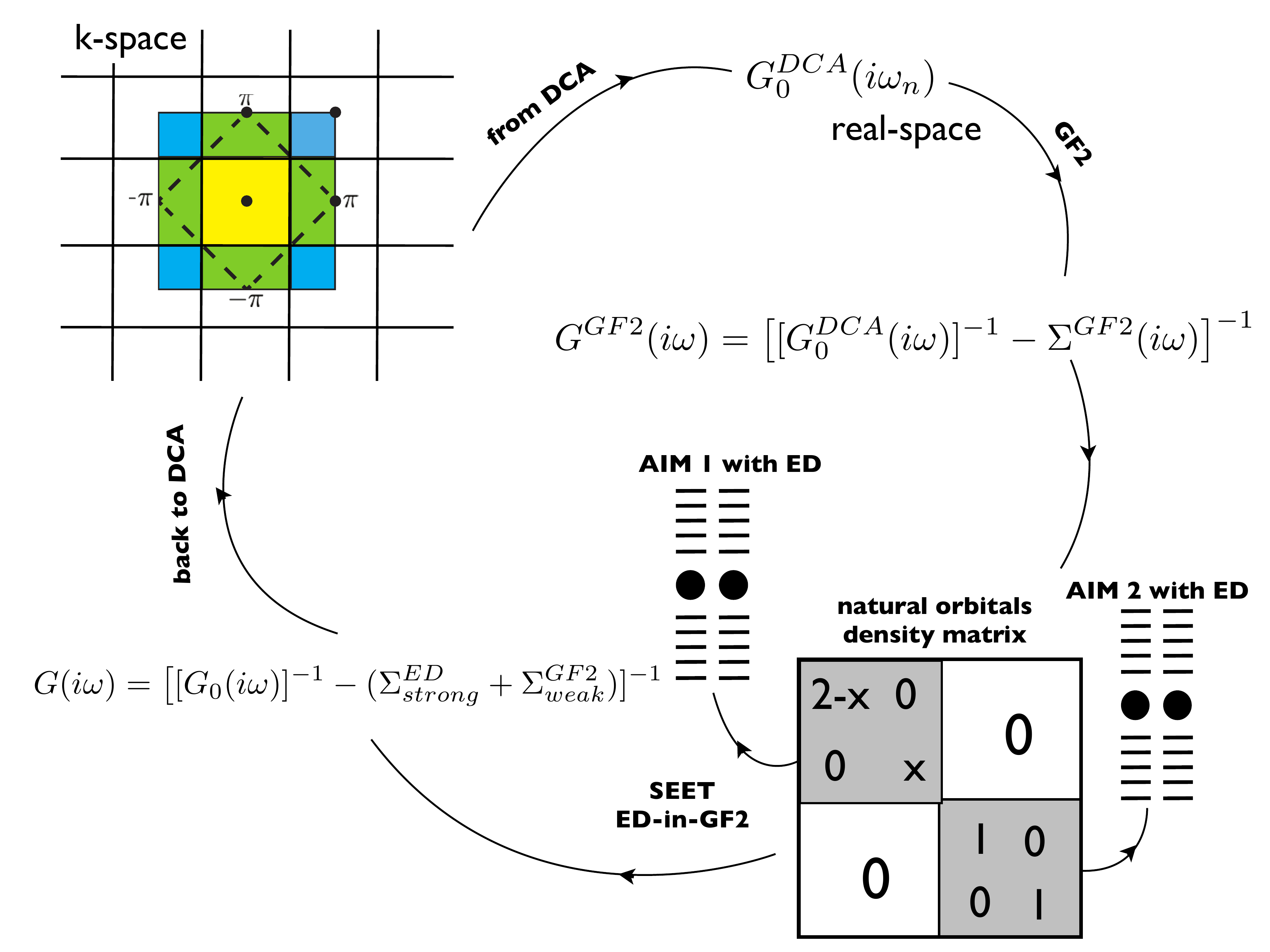}
\caption{\label{scheme} Schematic view of the DCA+ED-in-GF2 procedure used for treating the 4-site cluster DCA approximation to the 2D Hubbard model. Note that `strong' denotes quantities in the correlated subspace, and `weak' quantities defined on the entire system.}
\end{figure}

We calibrate SEET(ED-in-GF2) for the $2\times 2$ dynamical cluster approximation (DCA) to the 2D Hubbard model,\cite{Hettler00,Maier05} and consequently the Hamiltonian from Eq.~\ref{ham} is defined for $t$ describing nearest neighbor hopping only and $U$ exclusively on-site interactions. 
DCA provides the non-interacting Green's function (in Eq.~\ref{G0}) which is then employed to obtain the GF2 self-energy from Eq.~\ref{GF2}.  Subsequently, we construct the one-body density matrix and choose a pair of two-site impurities to be treated by ED. The occupations of the four site cluster in natural orbitals are 2-$x$, 1, 1, $x$, where for most regimes $x$ is not a small number, thus the orbitals with occupations 2-$x$, and $x$ are not any longer weakly correlated. This motivates us to choose two separate impurity problems with orbitals occupied as (1,1) and (2-$x$,$x$) and treat them as a pair of two-site impurities embedded into the GF2 description. This means that only the interactions between these two-site impurities are treated at the GF2 level. A schematic description of the DCA+ED-in-GF2 iterative scheme is shown in Fig.~\ref{scheme}. SEET allows us to treat multiple embedded impurities which is computationally advantageous since in realistic cases the number of strongly correlated orbitals may be too large for current solvers such as ED or the hybridization expansion.

Testing ED-in-GF2 using SEET on the DCA approximation to the 2D Hubbard model provides a worst case scenario for a multi-scale embedding scheme, since in multiple regimes the 4-site cluster does not display a separation of energy scales or any `weakly' and `strongly' correlated orbitals as typically found in realistic materials. Rather, in the Mott regime of the 2D Hubbard model,  all orbitals are strongly correlated, providing a  stringent test of the  SEET with ED-in-GF2 method.
\begin{figure} 
\centering
\includegraphics[width=1.0\columnwidth]{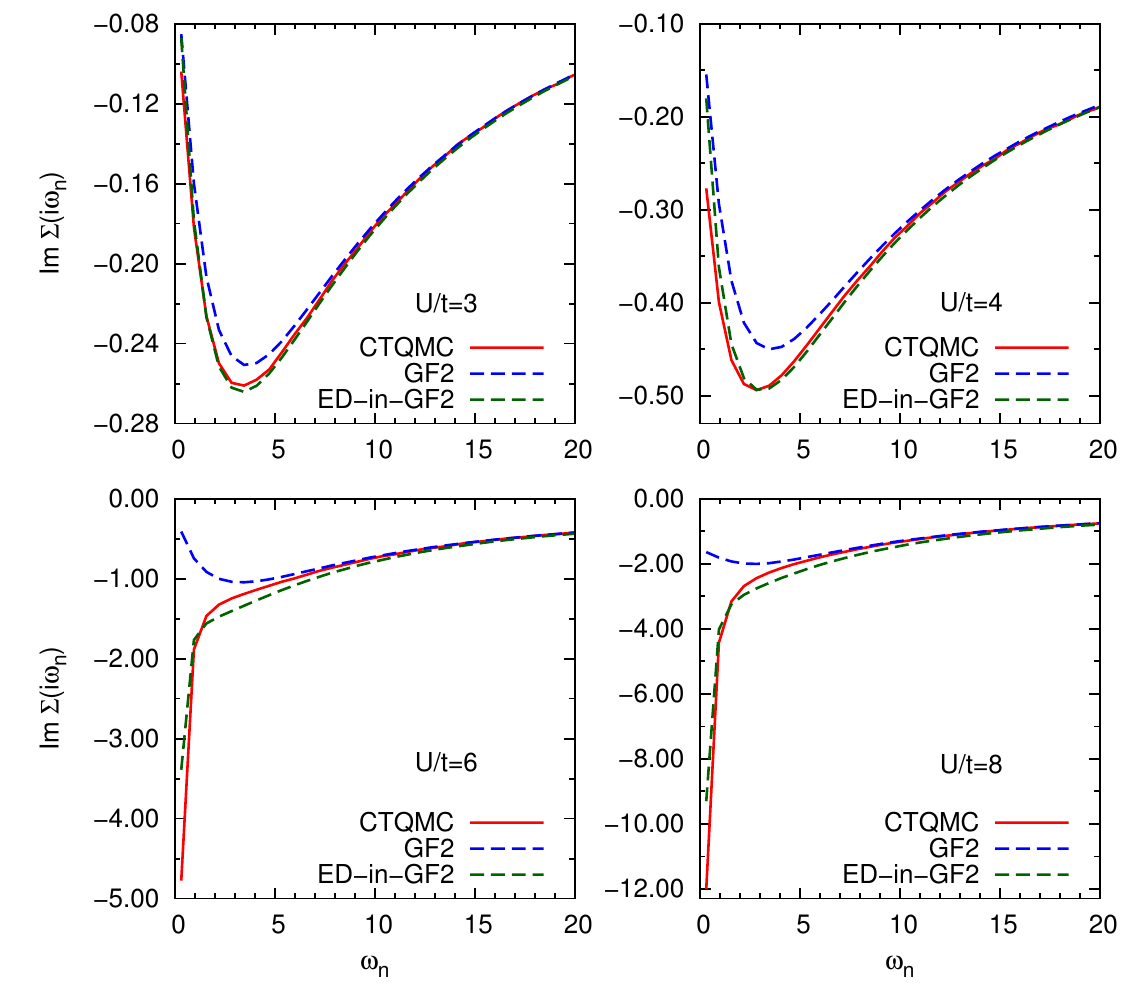}
\caption{\label{half_m} The imaginary part of the on-site CT-QMC, GF2 and ED-in-GF2 self-energy obtained for a 4-site cluster of the half-filled 2D Hubbard model for various regimes with $\beta=10t$.}
\end{figure}

In Fig.~\ref{half_m}, the imaginary part of the self-energy is plotted for the half-filled case. For weak coupling, {\it i.e.} $U/t <4,$ GF2 recovers the QMC results well. While for $U/t=3$ ED-in-GF2 corrects the  GF2 result only slightly,  for $U/t=4$, the improvement is more substantial. In this case, the ED-in-GF2 recovers QMC results and is a quantitative correction to the qualitatively correct GF2 curve.

As expected, in the Mott regime, $U/t=6$ and $8$, GF2 fails to recover the self-energy even qualitatively. Note that an IPT-like fitting of the large-$U$ limit to the atomic limit would be possible for this particular example but not in general, as it requires the determination of the local physics at exponential (in $n$) cost. In the Mott regime, ED-in-GF2 recovers to a decent quantitative accuracy the QMC self-energy for both $U/t=6$ and $8$.

\begin{figure}
\includegraphics[width=1.0\columnwidth]{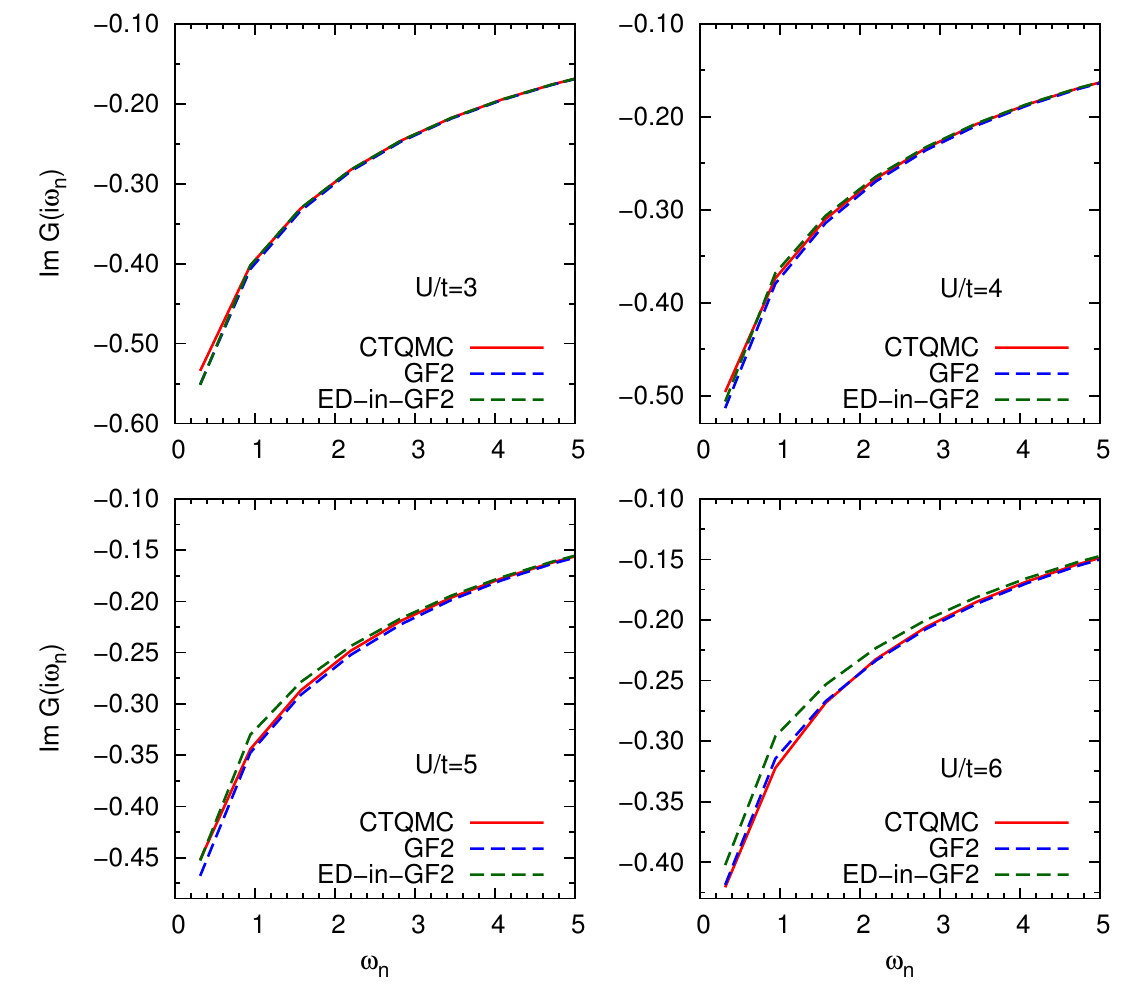}
\caption{\label{doped_img1} The imaginary part of the on-site CT-QMC, GF2 and ED-in-GF2 Matsubara Green's function obtained for a 4-site cluster of the 2D Hubbard model at $10\%$ doping, for $U/t=3, 4, 5,$ and $6$ with $\beta=10t$.}
\end{figure}

\begin{figure}
\includegraphics[width=1.0\columnwidth]{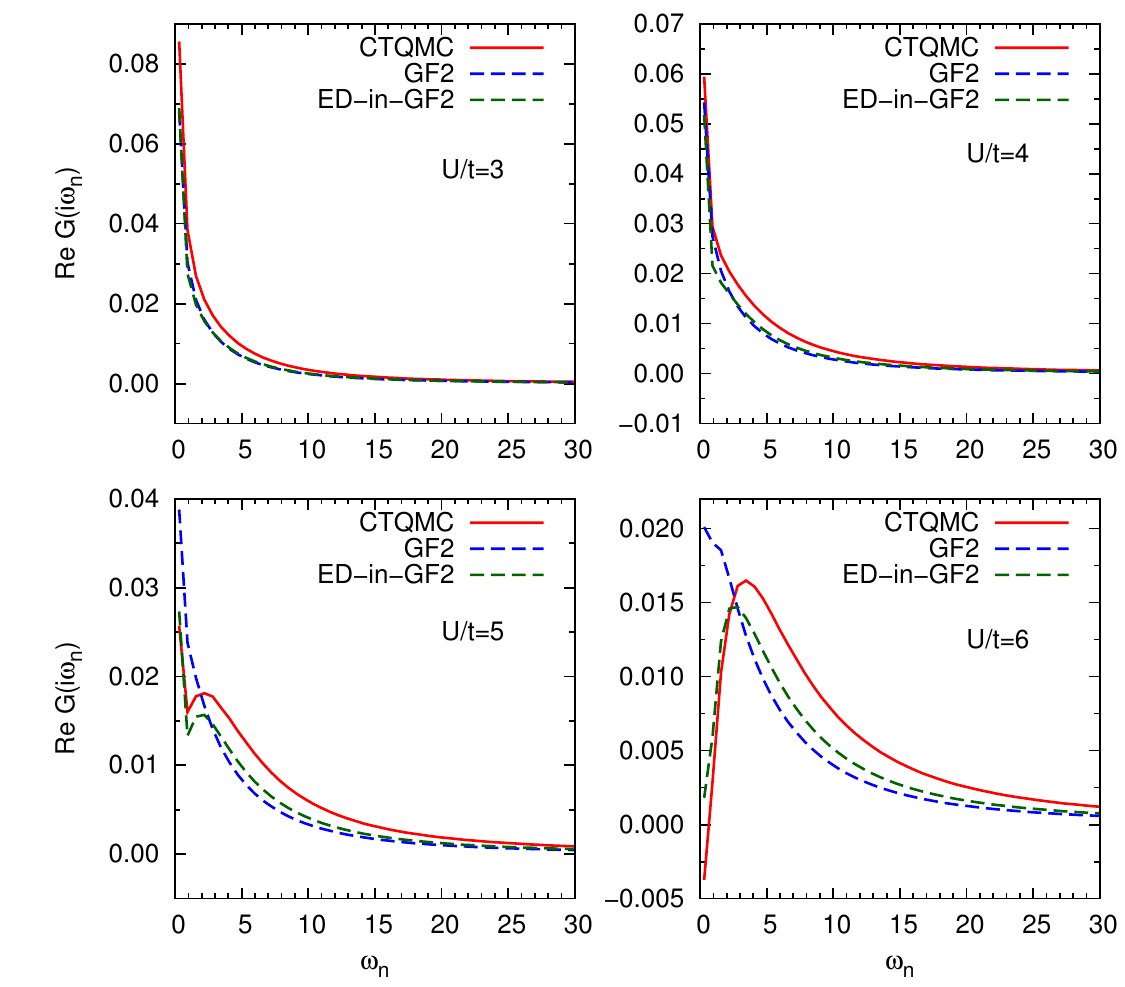}
\caption{\label{doped_real1} The real part of the on-site CT-QMC, GF2 and ED-in-GF2 Matsubara Green's function obtained for a 4-site cluster of the 2D Hubbard model at 10\% doping, for $U/t=3,4, 5,$ and $6$ with $\beta=10t$.}
\end{figure}

In Figs.~\ref{doped_img1} and~\ref{doped_real1}, we examine several interesting regimes at $10\%$ doping, where the system exhibits the behavior of a strongly correlated Fermi liquid. In these cases, we report real and imaginary parts of Green's functions rather than self-energies, since a slight difference in chemical potentials between different methods results in a shift of the Hartree term in the large-$\omega$ limit. 

The imaginary part of Green's function shows a good quantitative agreement between CT-QMC and ED-in-GF2 for multiple U/t regimes.
The real part of Green's function shows more differences than the imaginary part. In the weak coupling regime illustrated in Fig.~\ref{doped_real1}, for $U/t=3$ and $U/t=4$, all the QMC, GF2 and ED-in-GF2 real parts of Green's functions are close. 
The $U/t=5$ and $U/t=6$ regimes are more correlated and GF2 yields a qualitatively incorrect result. ED-in-GF2 corrects this result and provides a quantitative agreement with CT-QMC.
\section{Conclusions}\label{sec:conclusions}

We introduced a general self-energy embedding theory (SEET) for correlated systems and performed a comparison of SEET(ED-in-GF2) on a strongly correlated system for which the exact solution is known, the 4 site cluster DCA approximation to the 2D Hubbard model. This model has a continuous dispersion and shows a range of correlated phases, thus providing us with a detailed assessment of strengths and weaknesses of our method. However, it does not illustrate the effect of non-local interactions. Since in multiple regimes a clear separation of energy scales is not present, this model provides a rigorous test for a multi-scale method. We were able to show that ED-in-GF2 provides accurate results for the 4 site Hubbard model in the weakly correlated, intermediately correlated, and strongly correlated regimes, at and away from half-filling. While the solution in the strongly correlated embedded subset of orbitals has exponential scaling in our case, the total self-energy for the strongly correlated orbitals can be assembled using solutions of multiple small impurity problems. The calculation of the properties of the weakly coupled orbitals with GF2 scales as $O(N^5)$, making SEET(ED-in-GF2) an ideal tool for the simulation of realistic materials. Extensions using other diagrammatic or correlated methods, such as SEET(QMC-in-GF2) or methods based on GW are straightforward.

In real materials, the number of weakly correlated orbitals in the unit cell is significantly larger than the number of strongly correlated orbitals, thus providing an ideal situation where many orbitals can be treated cheaply by GF2 while the number of orbitals treated by ED remains small.  Moreover, the SEET(ED-in-GF2) hybrid is easy to implement and, since it does not use frequency dependent effective interactions, can be trivially extended to employ different solvers for the strongly correlated part, such as truncated CI variants with a suitably chosen active space or QMC hybridization expansions. Similarly, the weakly correlated part can be treated by different levels of perturbation theory or cheap truncated CI methods, instead of GF2.  Our ED-in-GF2 method can be adjusted to yield more accurate results, either by increasing the order of the perturbative treatment (e.g. by employing FLEX or GW), or by increasing the number of orbitals treated by ED. These limits therefore provide a rigorous assessment of the convergence of the self-energies. Since a set of strongly correlated orbitals in SEET is chosen based on a unique mathematical criterion,the method has the potential for becoming a black box method for realistic correlated materials calculations.

\acknowledgments{
D.Z. and A.K. acknowledge support by DOE ER16391, E.G. support from the Simons collaboration on the many-electron problem. We thank P. Werner for helpful comments on the first draft of this manuscript.
}

%

%\bibliography{short,pr1}
\end{document}